\documentclass[dvips,12pt,a4paper]{article}
\usepackage{epsfig}
\usepackage{a4wide}
\begin{document}
\pagestyle{empty}
\begin{titlepage}
\vspace{1. truecm}
\begin{center}
\begin{Large}
{\bf The description of multiplicity distributions\\
in high energy  $e^+e^-$ annihilation within\\
the framework of  Two Stage Model
\footnote
{Talk given at the International School
-Seminar "The Actual
Problems of Particle Physics", Gomel,  August 7 -- 16, 2001
}
}
\end{Large}

\vspace{2.0cm}
{\large  E.S. Kokoulina}
\\[0.3cm]
The Pavel Sukhoy Gomel State Technical University, 246746  Belarus

\end{center}
\vspace{1.5cm}

\begin{abstract}
\noindent
The multiplicity distributions in high energy $e^+e^-$~
annihilation are described within the Two Stage Model.
It is shown that oscillations in sign of the ratio
of factorial cumulant moments over factorial moments
of increasing order are confirmed.
Parameters of parton stage and hadronization
of this model are analyzed.

\noindent

\vspace*{3.0mm}

\noindent
\end{abstract}

\end{titlepage}

\pagestyle{plain}

\section{Introduction}
Multiparticle production (MP) is one of important
regions in high energy physics \cite{SIS}. Modern accelerators
had made possible the intensive and detailed study
of multiparticle processes. Quantum chromodynamics~(QCD),
developing theory of high energy physics \cite{POL}
and a lot of phenomenologic models are tested by the
process of MP. We should use phenomenologic models
on account of absent total theory of QCD.

MP is begun at high energy. Among of all
producing particles we can observe a lot of hadrons.
So, number of producing particles can be 50 and
more in $e^+e^-$ annihilation \cite{OPA}.

On one hand we want to know high energy
physics, but on the other hand the increase of the
inelastic channels is making difficulty in the
description of this process with customary methods by
using of the diagram technics. We have similarly
situation as we had at the time of developing
the thermodynamics and statistical physics.
The total analysis of MP process is very
difficult. The consideration of such reactions
begins from the analysis of behavior
of charge multiplicity.

{\bf Definition}: Multplicity is the number of
secondaries  {\bf n} in process MP:
\begin{equation}
\label{1}
{\bf A}  + {\bf B} \rightarrow {\bf a_1}+{\bf a_2}+
\ldots +{\bf a_n}
\end{equation}
{\bf Definition}: Multiplicity distribution (MD) $\bf P_n$
is the ratio cross-sections ${\sigma_n}$ to
$\sigma=\sum\limits_n\sigma_n$
$$
P_n=\sigma_n/\sigma.
$$
This quantity has such sense: the probability
of producing of $\bf n$ charge particles in
process (\ref{1}).

The description MP with using statistical
methods is based on a lot of particles in
this process. We can construct common
quantities like mean values, moments of
MD, correlations and so on.

The history of study MP is very interesting.
It connects on one hand with the developing
theory and on the other hand with increase
energy of accelerators.
We have stopped on the studing MP in $e^+e^-$ -
annihilation at high energy. At accordance
with theory of strong interactions QCD this
MP can be realize through the production of $\gamma$-- or $Z^0$--boson
into two quarks:
\begin{equation}
\label{2}
e^+e^-\rightarrow(Z^0/\gamma)\rightarrow q\bar q
\end{equation}

We can analyze the general features of
hadronic decays up to the highest available
center-of-mass~(c.m.) energies. In the last
years LEP produced $e^+e^-$~-collisions of
c.m. energies at 172, 183 and 189 GeV. The
charge particle multiplicity can be more
than 60 \cite{OPA}.

QCD can describe process of fission partons
(quarks and gluons) at high energy, because
the strong coupling $\alpha_s$ is
small at that energy. This stage can be called as
the stage of cascade. After fission, when
partons have not high energy, they must
change into hadrons, which we can observe.
On this stage we shouldn't apply QCD,
because $\alpha_s$ is not small.
Phenomenological models are used for
description of hadronization (transformation
partons (quarks and gluons) into hadrons).

It is expected on the stage quark-gluon
cascade perturbative QCD will be applied
\cite{{KUV},{DDT}}. Certain features
of the predictions at the parton level
are expected to be insensitive to
details of the hadronization mechanism.
They can be tested directly by using
hadron distributions \cite{HOY}.

The hadronization models are more
phenomenological and are built on the
experience gained from the study of low--$\bf{p_T}$
hadron collisions.
It is proposed to use Two Stage Model~(TSM)
\cite{TSM} for description MD and other
characteristics of $e^+e^-$ annihilation.
It is usually considered that the
producing of hadrons from partons is
universal one. We can use these MD for
calculation factorial and cumulant
factorial moments and oscillations of
their ratio and so on.

Investigation of MP led to discovery of
jets. Jets phenomena can be studied in
all processes, where  energetic partons are
produced. The most common ones are in $e^+e^-$
annihilation, deep inelastic scattering
of $e$, $\mu$ or $\nu$ on nucleons and hadron-
hadrons scattering, involving high-$\bf p_T$
particles in final state.

The $e^+e^-$--reaction is simple for analysis,
as the produced state is pure $q\overline q$.
It is usually difficult to determine the
quark species on event-by-event basis. The
experimental results are averaged over the
quark type. Because of confinement the
produced quark and gluons fragment into
jets of observable hadrons.
\section{Two Stage Model}
Parton spectra in QCD quark and gluon jets
have been studied by Konishi~K., Ukawa~A. and
Veneciano~G.\cite{KUV}. Working at the leading
logarithm approximation and avoiding IR
divergences by considering finite $x$, the
probabilistic nature of the problem has
been established \cite{KUV}.

At the study of MP at high energy we decided
to use idea A.~Giovannini~\cite{GIO} for
describing QCD jets as Markov branching
process. Giovannini proposed to interpret
the natural QCD evolution parameter as the
thickness of the jets and their development
through subnuclear matter as Markov process.
The Markov nature of elementary process
("gluon fission", "quark bremsstrahlung"  and
"quark-pair creation") leads to unsuspected
properties of QCD jets. The stochastic
approach gives clean complete solutions for
the parton MD. QCD evolution parameter is
\begin{equation}
\label{3}
\bf Y =\frac{1}{2\pi b}\ln[1+ab\ln(\frac{Q^2}{\mu^2})] ,
\end{equation}
where $2\pi b=\frac{1}{6}(11N_C-2N_f)$ for a theory
with $N_C$ colours and $N_f$ flavours.

Three elementary process contribute to the quark
or gluon distributions into QCD jets:\\
(1) gluon fission;\\
(2) quark bremsstrahlung;\\
(3) quark pair creation.

Let $A\Delta Y$ be the probability that
gluon in the infinitesimal interval $\Delta Y$
will convert into two gluons, $\widetilde{A}\Delta Y$
be the probability that  quark will radiate a gluon,
and $B\Delta Y$ be the probability that
a quark-antiquark pair will be created from
a gluon. $A, \widetilde A, B$ are assumed to be
{\bf Y}-independent constants and each individual
parton acts independently from others, always
with the same infinitesimal probability.

Definite the probability that $m_g$ gluons and
$m_q$ quarks will be transformed into
$n_g$~gluons and $n_q$~quarks over a jet of
thickness Y and call it $P_{m_g,m_q;n_g,n_q}(Y)$.
The probability generating function for a gluon
jet and quark jet will be, respectivily
\begin{equation}
\label{4}
G(u_g,u_q;Y)=\sum\limits_{n_g,n_q=0}^{\infty}
P_{1,0;n_g,n_q}(Y)u_g^{n_g}u_q^{n_q} ,
\end{equation}
\begin{equation}
\label{5}
Q(u_g,u_q;Y)=\sum\limits_{n_g,n_q=0}^{\infty}
P_{0,1;n_g,n_q}(Y)u_g^{n_g}u_q^{n_q} .
\end{equation}
Due to the independent action of the
individual partons it can be shown
through straightforward that
\begin{equation}
\label{6}
\sum\limits_{n_g,n_q=0}^{\infty}
P_{m_g,m_q;n_g,n_q}(Y)u_g^{n_g}u_q^{n_q}=
[G(u_g,u_q;Y)]^{m_g}[Q(u_g,u_q;Y)]^{m_q} .
\end{equation}
Eq.(\ref{6}) says that from a probabilistic
point of view the total parton population
($m_g$ gluons and $m_q$ quarks) evolving
through thickness {\bf Y} behaves as
($m_g+m_q$) independent parton population,
each with one initial quark or gluon.
Since, the process is homogeneous in {\bf Y}
the transition probabIlities obey Chapman-~
Kolmogorov equations:
\begin{equation}
\label{7}
P_{m_g,m_q;n_g,n_q}(Y+Y')=\sum\limits_{l_g,l_q=0}^{\infty}
P_{m_g,m_q;l_g,l_q}(Y)P_{l_g,l_q;n_g,n_q}(Y') .
\end{equation}

For the gluon jet one obtains
\begin{equation}
\label{8}
P_{0,1;n_g,n_q}(Y+Y')=\sum\limits_{l_g,l_q=0}^{\infty}
P_{1,0;l_g,l_q}(Y)P_{l_g,l_q;n_g,n_q}(Y')
\end{equation}
and for a quark jet
\begin{equation}
\label{9}
P_{0,1;n_g,n_q}(Y+Y')=\sum\limits_{l_g,l_q=0}^{\infty}
P_{0,1;l_g,l_q}(Y)P_{l_g,l_q;n_g,n_q}(Y') .
\end{equation}
Then (see eq.(\ref{4}),(\ref{5}))
\begin{equation}
\label{10}
G(u_g,u_q;Y+Y')=G[G(u_g,u_q;Y'),Q(u_g,u_q;Y');Y] ,
\end{equation}
\begin{equation}
\label{11}
Q(u_g,u_q;Y+Y')=Q[G(u_g,u_q;Y'),Q(u_g,u_q;Y');Y] .
\end{equation}
Making the substitution $Y'\to \Delta Y$, we obtain
\begin{equation}
\label{12}
\frac{\partial G(u_g,u_q;Y)}{\partial Y}=
\frac{\partial G}{\partial u_g}w^{(g)}(u_g,u_q)+
\frac{\partial G}{\partial u_q}w^{(q)}(u_g,u_q) ,
\end{equation}
\begin{equation}
\label{13}
\frac{\partial Q(u_g,u_q;Y)}{\partial Y}=
\frac{\partial Q}{\partial u_g}w^{(g)}(u_g,u_q)+
\frac{\partial Q}{\partial u_q}w^{(q)}(u_g,u_q) ,
\end{equation}
where the infinitesimal generating func\-tion
for qu\-ark and glu\-on jets \\
($w^{(g)}(u_g,u_q)$ and $w^{(q)}(u_g,u_q)$, respectively) are
\begin{equation}
\label{14}
w^{(g)}(u_g,u_q)=\sum\limits_{m_g,m_q=0}^{\infty}
a^{(g)}_{m_g,m_q}u_g^{m_g}u_q^{m_q}=
(-A-B)u_g+Au_g^2+Bu^2_q,
\end{equation}
\begin{equation}
\label{15}
w^{(q)}(u_g,u_q)=\sum\limits_{m_g,m_q=0}^{\infty}
a^{(q)}_{m_g,m_q}u_g^{m_g}u_q^{m_q}=
(-\widetilde Au_q+\widetilde Au_qu_g).
\end{equation}

Eqs. (\ref{12}) and (\ref{13}) recognize the
forward Kolmogorov equations for the generation
functions of the transition probabilities
$P_{m_g,m_q;n_g,n_q}(Y)$.
The corresponding backward Kolmogorov
equations follow from eqs.(\ref{10},\ref{11})
\begin{equation}
\label{16}
\frac{\partial G}{\partial Y}=
w^{(g)}[G(u_g,u_q;Y),Q(u_g,u_q;Y)],
\end{equation}
\begin{equation}
\label{17}
\frac{\partial Q}{\partial Y}=
w^{(q)}[G(u_g,u_q;Y),Q(u_g,u_q;Y)].
\end{equation}
Recalling eqs.(\ref{14})and (\ref{15}),
eqs. (\ref{16}) and (\ref{17}) become
\begin{equation}
\label{18}
\frac{\partial G}{\partial Y}=
-AG+AG^2-BG+BQ^2,
\end{equation}
\begin{equation}
\label{19}
\frac{\partial Q}{\partial Y}=
-\widetilde AQ+\widetilde AQG.
\end{equation}

These eqs. were obtained in \cite{GIO}.
We can find  the probability for quark
(or gluon) to produce in the interval
($Y+\Delta Y$) $n_g$ gluons and $n_q$
quarks through process (1)-(3).
It follows for gluon jet
$$
P_{1,0;n_g,n_q}(Y+\Delta Y)=
[1-\widetilde An_q\Delta Y-An_q\Delta Y-
Bn_g\Delta Y]P_{1,0;n_g,n_q}(Y)
$$
$$
+\widetilde An_q\Delta YP_{1,0;n_g-1,n_q}+
A(n_g-1)\Delta YP_{1,0;n_g-1,n_q}(Y)
$$
\begin{equation}
\label{20}
+B(n_g+1)\Delta YP_{1,0;n_g+1,n_q-2}(Y)+
o(\Delta Y)
\end{equation}
Letting $\Delta Y\to 0$ we get the system
of differential equations:
$$
\frac{dP_{1,0;n_g,n_q}(Y)}{dY}=
[\widetilde An_q-Bn_g-An_g]P_{1,0;n_g,n_q}
$$
$$
+\bar An_gP_{1,0;n_g-1,n_q}(Y)+
A(n_g-1)P_{1,0;n_g-1,n_q}(Y)
$$
\begin{equation}
\label{21}
+B(n_g+1)P_{1,0;n_g+1,n_q-2}(Y).
\end{equation}

Finding the explicit solutions both
in terms of generating functions
(\ref{18}),(\ref{19}) or of the
exclusive cross-sections(\ref{20})
is not easy. Approximate solutions
can be obtained for particular case
$B=0,\quad \widetilde A \neq A\neq 0$.

The meaning of this  approximation
is next: we don't allow gluons to
split into quark-antiquark pair.
Eqs. for gluon and quark jets
in this case are
\begin{equation}
\label{22}
\frac{dP_{1,0;n_g,0}(Y)}{dY}=A(n_g-1)
P_{1,0;n_g-1,0}(Y)-An_gP_{1,0;n_g,0}(Y),
\end{equation}
$$
\frac{dP_{0,1;n_g,1}(Y)}{dY}=
-\bar AP_{0,1;n_g,1}(Y)-An_gP_{0,1;n_g,1}(Y)
$$
\begin{equation}
\label{23}
+\bar AP_{0,1;n_g-1,1}(Y)+An_gP_{0,1;n_g,1}(Y)
\end{equation}
with initial conditions $P_{0,1;0,1}(0)=1$,
\quad $P_{1,0;1,0}(0)=1$,\quad
$P_{1,0;n_g,0}(0)=0$ and $P_{0,1;n_g,1}(0)=0$\quad
$\forall n_g >1$.

For gluon jet we have
\begin{equation}
\label{24}
P_{1,0;n_g,0}(Y)=\frac{1}{<n_g>}
\left(1-\frac{1}{<n_g>}\right)^{n_g-1}
\end{equation}
with a average gluon multiplicity
$<n_g>=e^{AY}$ .
For quark jet (\ref{23}) one obtains MD
\begin{equation}
\label{25}
P_{0,1;0,1}(Y)=e^{-\widetilde AY},
\end{equation}
\begin{equation}
\label{26}
P_{0,1;n_g,1}(Y)=\frac{\mu(\mu+1)\dots
(\mu+n_g-1)}{n_g!}e^{-\widetilde AY}
(1-e^{-AY})^{n_g},
\end{equation}
where $\mu=\frac{\widetilde A}{A}$ .
The average gluon multiplicity will be
$<n_g>=\mu (e^{AY}-1)$ and the
normalized exclusive cross section
for producing $n_g$ gluons
$$
\frac{\sigma_{n_g}^{(0,q)}}{\sigma_{tot}}\equiv
P_{0,1;n_g,1}(Y)=
$$
\begin{equation}
\label{27}
=\frac{\mu(\mu+1)\dots(\mu+n_g-1)}{n_g!}
\left[\frac{<n_g>}{<n_g>+\mu}\right]^{n_g}
\left[\frac{\mu}{<n_g>+\mu}\right]^{\mu}.
\end{equation}

The generating function will be given by
\begin{equation}
\label{28}
Q=\sum\limits_{n_g=0}^{\infty}
u_g^{n_g}u_qP_{0,1;n_g,1}(Y)=
u_q\left[\frac{e^{-AY}}{1-u_g(1-e^{-AY})}
\right]^{\mu}.
\end{equation}
Eq.(\ref{27}) is Polya-Egenberger
distribution, where $\mu$ is not integer.

In Two Stage Model \cite{TSM} we took
(\ref{28}) for description of parton
stage and added for final stage
hadronization supernarrow distribution of
Bernulli (binomial distribution).
We chose it based oneself on analysis
experimental data in $e^+e^-$-
annihilation at 9 - 22 GeV. Second
correlation moments were negative
at this energy and it was one,
which could describe experiment.

We  consider  that  hypothesis
of soft  bleachment is  right.
We added stage of hadronization to
parton stage with aid of their
factorization. MD in this
process can be writing
\begin{equation}
\label{29}
P_n(s)=\sum\limits_mP^P_mP_n^H(m,s),
\end{equation}
where $P_m^P$ - MD for partons
(\ref{27}), $P_n^H(m,s)$
- MD for hadrons, are produced from
$m$ partons on the stage of
hadronization.

In accordance with TSM the stage
of hard fission of partons
is described by negative binomial
distribution (NBD) for quark jet
\begin{equation}
\label{30}
P_m^P(s)=\frac{k_p(k_p+1)\dots(k_p+
m-1)}{m!}\left(\frac{\overline m}
{\overline m+k_p}\right)^{k_p}\left(
\frac{k_p}{k_p+\overline m}\right)^m,
\end{equation}
where $k_p=\widetilde A/A$, {\quad} $\overline m=\sum\limits_mmP_m^P$.
We are neglecting process (3) quark pair production ($B=0$). Two quarks
fragment to parton independently  each other. Total MD of two quarks
is equal to (\ref{30}) too. MD $P_m^P$ and generating function
for MD $Q^P(s,z)$ are
\begin{equation}
\label{31}
P_m^P=\frac{1}{m!}\frac{\partial^m}
{\partial z^m}\left.Q^P(s,z)\right|_{z=0},
\end{equation}
\begin{equation}
\label{32}
Q_m^P(s,z)=\left[1+\frac{\overline m}
{k_p}(1-z)\right]^{-k_p}.
\end{equation}
From TSM MD of hadron
are formed from parton are described
in form
\begin{equation}
\label{33}
P_n^H=C^n_{N_p}\left(\frac{\overline n^h_p}
{N_p}\right)^n\left(1-\frac{\overline n_p^h}
{N_p}\right)^{N_p-n},
\end{equation}
($C_{N_p}^n$ - binomial coefficient) with
generating function
\begin{equation}
\label{34}
Q^h_p=\left[1+\frac{\overline n^h_p}
{N_p}(z-1)\right]^{N_p},
\end{equation}
where $\overline n^h_p$ and $N_p$
($p=q,g$) have sense of average
multiplicity and maximum secondaries
of hadrons are formed from parton
on the stage of hadronization.
MD of hadrons in $e^+e^-$
annihilation are determined by
convolution of two stages
\begin{equation}
\label{35}
P_n(s)=\sum\limits_{m=o}^{\infty}
P_m^P\frac{\partial^n}{\partial z^n}
(Q^H)^{2+m}|_{z=0}
\end{equation}

Further we assume next simplification
for second stage: $\frac{\overline n^h_q}
{N_q}\approx\frac{\overline n^h_g}
{N_g}$ . Let $\alpha=\frac{N_g}{N_q}$,
$N=N_q$, $\overline n^h=\overline n^h_q$.
Then
$$
Q^H_q=\left(1+\frac{\overline n^h}
{N}(z-1)\right)^N,
$$
$$Q^H_g=\left(1+\frac{\overline n^h}
{N}((z-1)\right)^{\alpha N}.
$$
From(\ref{35}) we have
\begin{equation}
\label{36}
P_n(s)=\sum\limits_{m=0}P_m^P
C^n_{2+\alpha m)N}\left(\frac{\overline n^h}
{N}\right)^n\left(1-\frac{\overline n^h}
{N}\right)^{(2+\alpha m)N-n}.
\end{equation}
For comparing with experimental
data the normalized factor $\Omega$ was
introduced into (\ref{36}) and a number
of summarized gluons were restricted
by $M_g$ - maximal
number of gluons are giving cascade
\begin{equation}
\label{37}
P_n(s)=\Omega \sum\limits_{m=0}
^{M_g}P_m^PC^n_{(2+\alpha m)N}
\left(\frac{\overline n^h}
{N}\right)^n\left(1-\frac
{\overline n^h}{N}\right)^{(2
+\alpha m)N-n}.
\end{equation}

The results of comparison of
model expression MD(\ref{37})
with experimental data \cite{OPA},
\cite{DAT} represented in the Table 1.
Apparently that behavior of parameters
of TSM is giving the reasonable
picture of $e^+e^-$ annihilation
process. The average
multiplicity of gluons $\overline m_g$
formed on fission stage
changes from 3 - 4 at 22 GeV to
15 - 16 at 189 Gev. This increase
takes place with the growing energy,
but at the some energy this value
accepts a big one (at 43.6 GeV).

From QCD is following that the
parameter $k_p$ must
decrease at more high energies. The
parameter of parton spectra $k_p$
changes: from 140 at low energy
(22 GeV) up 5 - 6 at 172 - 189~GeV.
QCD are giving limiting value $k_p$
is equal to 1.

The most interesting dependencies
are discovered for parameters of
hadronization $\overline n_q^h$,
$N_q$  and $\alpha$ . The parameter
$N_q$ determines maximum number of
hadrons, which can
be formed  from
quark on stage of
hadronization.
It changes unusually, it can be 5 - 6
or 8 - 15 at different energies, so
it equal about 9 at 172 GeV, but
equal 54.6 at 183 and again 11 at
189 GeV. After that analysis we
can say about the existing some
special energies when process
hadronization changes (at
those energies value of $\chi ^2$
is changed unconsiderably).

At the same time the parameter
of hadronization
$\overline n_q^h$ (mean number
of hadrons formed from quark
on stage of hadronization)
discovers insignificant growth
with increasing of energy. It
is changed from 3 - 4 at 22~GeV
up 4 - 5 at 172 - 189 GeV. The
existing models give quantity
$n_q^h$ equal two (it is
observed really at energies
lower than 40 GeV). Such
behavior of parameter may be
connected with the growth
of spectrum of mass hadron states
(appearance of new mass states
with increase of energy). It
is interesting that the ratio
$\frac{\overline n^h_q}{N_q}$
(the parameter of PBD)
is approximately regularity
and is equal to $1/2$.
It should mark  breaking
such behavior at some energies
(43.6, 183 GeV) again.

Parameter $\alpha$ is stayed
approximately constant and
equal to 0.2 in order case
and become smaller than 0.1
at special cases (43.6 and
183 GeV). The number of
hadrons are produced from
quark jet on the stage of
hadronization is  more
than from gluon jet. The
normalized factor $\Omega$
is stayed constant and equal to 2.
\section{Oscillation of moments in MD}
At the recent years it was shown
\cite{FAC} that the ratio of
factorial cumulative moments over
factorial moments changes sign as
a function of order. We can use
MD formed in TSM for explanation of
this phenomena.

\input{epsf}
\begin{figure}
\vspace{40mm}
\begin{tabular}{c}
\begin{picture}(100,160)
\put(-20,-40){
\epsfxsize=7cm
\epsfysize=7cm
\epsfbox{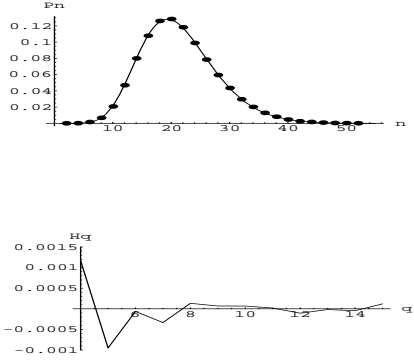} }
\end{picture}
\end{tabular}
\caption{\protect\it
Multiplicity distributions $P_n$ and $H_q$
as a function of $q$ at $Z^0$.
}
\label{1f}
\end{figure}

The factorial moments can be obtained
from MD $P_n$ through the relation
\begin{equation}
\label{38}
F_q=\sum\limits_{n=q}^{\infty}
n(n-1)\dots(n-q+1)P_n ,
\end{equation}
and factorial cumulative moments
are found from expression
\begin{equation}
\label{39}
K_q=F_q-\sum\limits_{i=1}^{q-1}C_{q-i}^{i}
K_{q-i}F_i .
\end{equation}
The ratio of their quantities is
\begin{equation}
\label{40}
H_q=K_q/F_q.
\end{equation}
We can use the generating function for
MD of hadrons (\ref{36}) in $e^+e^-$
annihilation G(z)
$$
G(z)=\sum\limits_{m=0}P_m^g[Q_g^H(z)]^m Q^2_q(z)=
$$
\begin{equation}
\label{41}
=Q^g (Q^H_g(z))^m Q^2_q(z) .
\end{equation}
We are calculating $F_q$ and $K_q$
in TSM using (\ref{41})
\begin{equation}
\label{42}
F_q=\frac{1}{\overline n^q(s)}\left.\frac
{\partial^qG}{\partial z^q}\right|_{z=0}
\end{equation}
\begin{equation}
\label{43}
K_q=\frac{1}{\overline n^q(s)}\left.\frac
{\partial ^q\ln G}{\partial z^q}\right|_{z=0}.
\end{equation}
The expression(\ref{41}) for G(z)
after taking a logarithm
$$
\ln G(s,z)=-k_p\ln [1+\frac{\overline m}{k_p}(1-Q^H_g)]+2\ln Q^H_q
$$
and the expansion to series in power
on $Q^H_g$ will be
\begin{equation}
\label{44}
\ln G(s,z)=k_p\sum\limits_{m=1}\left(
\frac{\overline m}{\overline m+k_p}\right)^m
\frac{Q^m_g}{m}+2\ln Q .
\end{equation}
Inserting $Q_g$ into (\ref{44})
$$\ln G(s,z)=k_p\sum\limits_{m=0}\left(
\frac{\overline m}{\overline m+k_p}
\right)^m\frac{1}{m}\left[1+\frac{\overline n^h}{N}
(z-1)\right]^{\alpha mN}+
$$
$$
+2N\ln [1+\frac{\overline n^h}{N}(z-1)],
$$
and using (\ref{43}) we obtain
$$
K_q=\left(k_p\sum_{m=1}\alpha m (\alpha m -\frac{1}{N})\dots
(\alpha m-\frac{q-1}{N})\left(\frac{\overline m}
{\overline m+k_p}\right)^m\frac{1}{m}\right.
$$
\begin{equation}
\label{45}
\left.-2(-1)^q\frac{(q-1)!}{N^{q-1}}\right)\left(
\frac{\overline n^h}{\overline n(s)}\right)^q
\end{equation}
where $\overline n(s)$ is the average
multiplicity hadrons in process (\ref{2}).
It is possible to find $F_q$ using (\ref{42})
\begin{equation}
\label{46}
F_q=\sum\limits_{m=0}(2+\alpha m)(2+\alpha m-
\frac{1}{N})\dots(2+\alpha m-\frac{q-1}{N})
P_m\left(\frac{\overline n^h}{\overline n(s)}
\right)^q
\end{equation}
with $P_m$ equal (\ref{30}).

The sought-for expression for $H_q$ will be
\begin{equation}
\label{47}
H_q=\Omega_1\frac{\sum\limits_{m=1}k_p\alpha m
(\alpha m-\frac{1}{N})\dots(\alpha m-\frac{q-1}
{N})(\frac{\overline m}{\overline m+k_p})^m-
2(-1)^q\frac{(q-1)!}{N^{q-1}}}
{\sum\limits_{m=0}(2+\alpha m)(2+\alpha m-
\frac{1}{N})\dots(2+\alpha m-\frac{q-1}{N})P_m}
\end{equation}
where $\Omega_1$ is the normalized factor.
The comparison with experimental data \cite{DAT}
shows that (\ref{47}) is describing the ratio of
factorial moments. It is seen minimum at q=5.

The immediate calculations $H_q$ based on
(\ref{38})-(\ref{40}) with using MD(\ref{37})
are giving very good description of the
oscillation value of $H_q$ ($\chi^2 \approx2$).
Significant oscillations begin near region
producing of $Z^0$ and can be explain by not
integer values of parameters of hadronization
$N_q$ and $N_g=\alpha N_q$ or by convolution
of wide NBD and narrow PBD.
\section{Conclusions}
It is shown that TSM does not contradict to
the experimental data on MD and the oscillations
ratio of factorial moments (small $\chi^2$)
(Figure 1).

\begin{center}
Table 1. Parameters of TSM.
\end{center}
\renewcommand{\tablename}{Table}
\begin{center}
\begin{tabular}{||c||c|c|c|c|c|c||}
\hline
\hline
$\sqrt s$ GeV &$\overline m$ &$k_p$&$N$&$\overline n^h$&$\alpha $&$\Omega $\\
\hline
\hline
14  &  .08 &$2.4* 10^8$& 27.7 & 2.87 & .97 &2\\
\hline
22  & 3.01 & 4.91      & 20.2 & 4.34 & .2  &2\\
\hline
34.8& 6.58 & 6.96      & 12.5 & 4.1  & .195&2\\
\hline
43.6& 10.9 & 39.4      &  6.1 & 2.43 & .386&2\\
\hline
91.4& 10.9 &  7.86     & 11.2 & 4.8  & .226&2\\
\hline
172 & 20.1 &  9.11     &  9.17& 4.34 & .195&1.98\\
\hline
183 & 13.2 &  1.48     & 54.6 & 8.9  & .086&2.06\\
\hline
189 & 15.1 &  6.9      & 11.6 & 5.15 & .215&2.01\\
\hline
\hline
\end{tabular}
\end{center}

\vspace*{20mm}
\setcounter{footnote}{0}
\end{document}